\documentclass[%
  reprint,
  superscriptaddress,
  amsmath,amssymb,
  aps,
  pra,
]{revtex4-2}
\usepackage{placeins}
\usepackage[english]{babel}
\DeclareUnicodeCharacter{2212}{\textminus}
\usepackage[utf8]{inputenc}
\usepackage{textgreek}
\usepackage{lipsum}

\usepackage{graphicx}
\graphicspath{{figs/}}          

\usepackage{bm}                 
\usepackage{upgreek}
\usepackage{braket}
\usepackage{siunitx}
\DeclareSIUnit\bar{bar}
\DeclareSIUnit\planck{h}

\usepackage{multirow}
\usepackage{dcolumn}

\usepackage{blindtext}          
\usepackage{changes}            
\usepackage{xcolor}

\usepackage[title]{appendix}
\usepackage{natbib}
\usepackage[colorlinks]{hyperref}
\usepackage[capitalise,nameinlink]{cleveref}
\crefname{section}{App.}{Apps.}

\definecolor{greengray}{RGB}{40,180,160}
\definecolor{navygray}{RGB}{110,140,170}

\newcommand{\crefadd}[2]{\hyperref[#1]{\cref{#1}#2}}
\newcommand{\Crefadd}[2]{\hyperref[#1]{\Cref{#1}#2}}
\hypersetup{
  citecolor={navygray},
  linkcolor={navygray},
  urlcolor={navygray},
}
\makeatletter
\newcommand*{\balancecolsandclearpage}{%
  \close@column@grid
  \cleardoublepage
  \twocolumngrid
}
\makeatother

\begin{document}
\title{Quantum Coherence in Superconducting Vortex States}

\author{Ameya~Nambisan}
\affiliation{IQMT,~Karlsruhe~Institute~of~Technology,~76131~Karlsruhe,~Germany}

\author{Simon~Günzler}
\affiliation{IQMT,~Karlsruhe~Institute~of~Technology,~76131~Karlsruhe,~Germany}
\affiliation{PHI,~Karlsruhe~Institute~of~Technology,~76131~Karlsruhe,~Germany}

\author{Dennis~Rieger}
\affiliation{IQMT,~Karlsruhe~Institute~of~Technology,~76131~Karlsruhe,~Germany}
\affiliation{PHI,~Karlsruhe~Institute~of~Technology,~76131~Karlsruhe,~Germany}

\author{Nicolas~Gosling}
\affiliation{IQMT,~Karlsruhe~Institute~of~Technology,~76131~Karlsruhe,~Germany}

\author{Simon~Geisert}
\affiliation{IQMT,~Karlsruhe~Institute~of~Technology,~76131~Karlsruhe,~Germany}

\author{Victor~Carpentier}
\affiliation{IQMT,~Karlsruhe~Institute~of~Technology,~76131~Karlsruhe,~Germany}

\author{Nicolas~Zapata}
\affiliation{IQMT,~Karlsruhe~Institute~of~Technology,~76131~Karlsruhe,~Germany}

\author{Mitchell~Field}
\affiliation{IQMT,~Karlsruhe~Institute~of~Technology,~76131~Karlsruhe,~Germany}

\author{Milorad~V.~Milošević}
\affiliation{Department~of~Physics, University~of~Antwerp,~B-2020~Antwerp,~Belgium}

\author{Carlos~A.~Diaz Lopez}
\affiliation{Institute~for~Complex~Quantum~Systems~and~IQST,~University~of~Ulm,~89069~Ulm,~Germany}

\author{Ciprian~Padurariu}
\affiliation{Institute~for~Complex~Quantum~Systems~and~IQST,~University~of~Ulm,~89069~Ulm,~Germany}

\author{Bj\"orn~Kubala}
\affiliation{German~Aerospace~Center~(DLR),~Institute of~Quantum~Technologies,~89081~Ulm,~Germany}
\affiliation{Institute~for~Complex~Quantum~Systems~and~IQST,~University~of~Ulm,~89069~Ulm,~Germany}

\author{Joachim~Ankerhold}
\affiliation{Institute~for~Complex~Quantum~Systems~and~IQST,~University~of~Ulm,~89069~Ulm,~Germany}

\author{Wolfgang~Wernsdorfer}
\affiliation{IQMT,~Karlsruhe~Institute~of~Technology,~76131~Karlsruhe,~Germany}
\affiliation{PHI,~Karlsruhe~Institute~of~Technology,~76131~Karlsruhe,~Germany}

\author{Martin~Spiecker}
\affiliation{IQMT,~Karlsruhe~Institute~of~Technology,~76131~Karlsruhe,~Germany}
\affiliation{PHI,~Karlsruhe~Institute~of~Technology,~76131~Karlsruhe,~Germany}

\author{Ioan~M.~Pop}
\email{ioan.pop@kit.edu}
\affiliation{IQMT,~Karlsruhe~Institute~of~Technology,~76131~Karlsruhe,~Germany}
\affiliation{PHI,~Karlsruhe~Institute~of~Technology,~76131~Karlsruhe,~Germany}
\affiliation{Physics~Institute~1,~Stuttgart~University,~70569~Stuttgart,~Germany}

\date{\today}
\begin{abstract}
Abrikosov vortices, where the superconducting gap is completely suppressed in the core, are dissipative, semi-classical entities that impact applications from high-current-density wires to superconducting quantum devices. In contrast, we present evidence that vortices trapped in granular superconducting films can behave as two-level systems, exhibiting microsecond-range quantum coherence and energy relaxation times that reach fractions of a millisecond. 
These findings support recent theoretical modeling of superconductors with granularity on the scale of the coherence length as tunnel junction networks, resulting in gapped vortices~\cite{kiselov2023gapful}.
Using the tools of circuit quantum electrodynamics, we perform coherent manipulation and quantum non-demolition readout of vortex states in granular aluminum microwave resonators, heralding new directions for quantum information processing, materials characterization, and sensing.
\end{abstract}
\maketitle

From the moment of its discovery, the antagonistic relation between superconductivity and magnetic field has provided a complex playground for experimentalists and theorists alike. 
The measurement of the critical field and the Meissner effect~\cite{Meissner_Ochsenfeld_1933} have anchored phase transition theories~\cite{London1935, ginzburg_theory_2009}, and the trapping of quantized flux inside superconductors has provided direct evidence for the existence of Cooper pairs~\cite{Deaver_Fairbank_1961, Doll_Näbauer_1961}. 
A hallmark of type II superconductivity in magnetic field is the formation of Abrikosov vortices: regions of locally suppressed gap that interact to form lattices~\cite{Abrikosov:1956sx}. 
Vortex dynamics is detrimental for a wide range of applications~\cite{tinkham2004introduction}, causing heating, flux noise, and magnetic hysteresis. 
On the other hand, pinned vortices enable quasiparticle trapping in their core, which enhances the critical current~\cite{Hebard_1977} of superconducting films, improves micro-cooler efficiency~\cite{peltonen2011magnetic}, boosts resonator quality factors~\cite{nsanzineza2014trapping}, and improves qubit coherence~\cite{wang2014measurement, vool2014non}. 
In all these cases, due to the normal state core, vortices can be understood within semi-classical models. 

Gap suppression in the vortex core stems from the crowding of supercurrent at its center, a consequence of continuity in the superconducting medium. 
Recent work~\cite{kiselov2023gapful} has proposed that in discretized systems, such as granular superconductors where non-superconducting regions separate superconducting islands, the vortex core can remain gapped and dissipationless. 
While quantum behavior has been revealed by tunneling of vortices in long Josephson junctions~\cite{Wallraff2003Sep} and thin films~\cite{Fruchter1991Apr, Liu_1992}, or via the zero-point motion of pinned vortices~\cite{Dutta_2021}, a direct measurement of coherence in superconducting vortex states remains elusive.

Here, we demonstrate that vortices trapped in a superconducting granular aluminum (grAl) resonator exhibit remarkably low loss and behave as effective spins. They can therefore be regarded as quantum bits (qubits) and have coherence times in the ${\si{\micro\second}}$ range.
These vortex qubit (VQ) states can be modeled by a double-well potential formed between pinning sites, modulated by their Gibbs energy in magnetic field.
We find VQs strongly coupled to the resonator and stable for weeks, allowing coherent control and quantum non-demolishing readout within the framework of circuit quantum electrodynamics~\cite{Blais__Review_cQED__2021}.

\begin{figure*}[htbp]
    \includegraphics[width=0.95\textwidth]{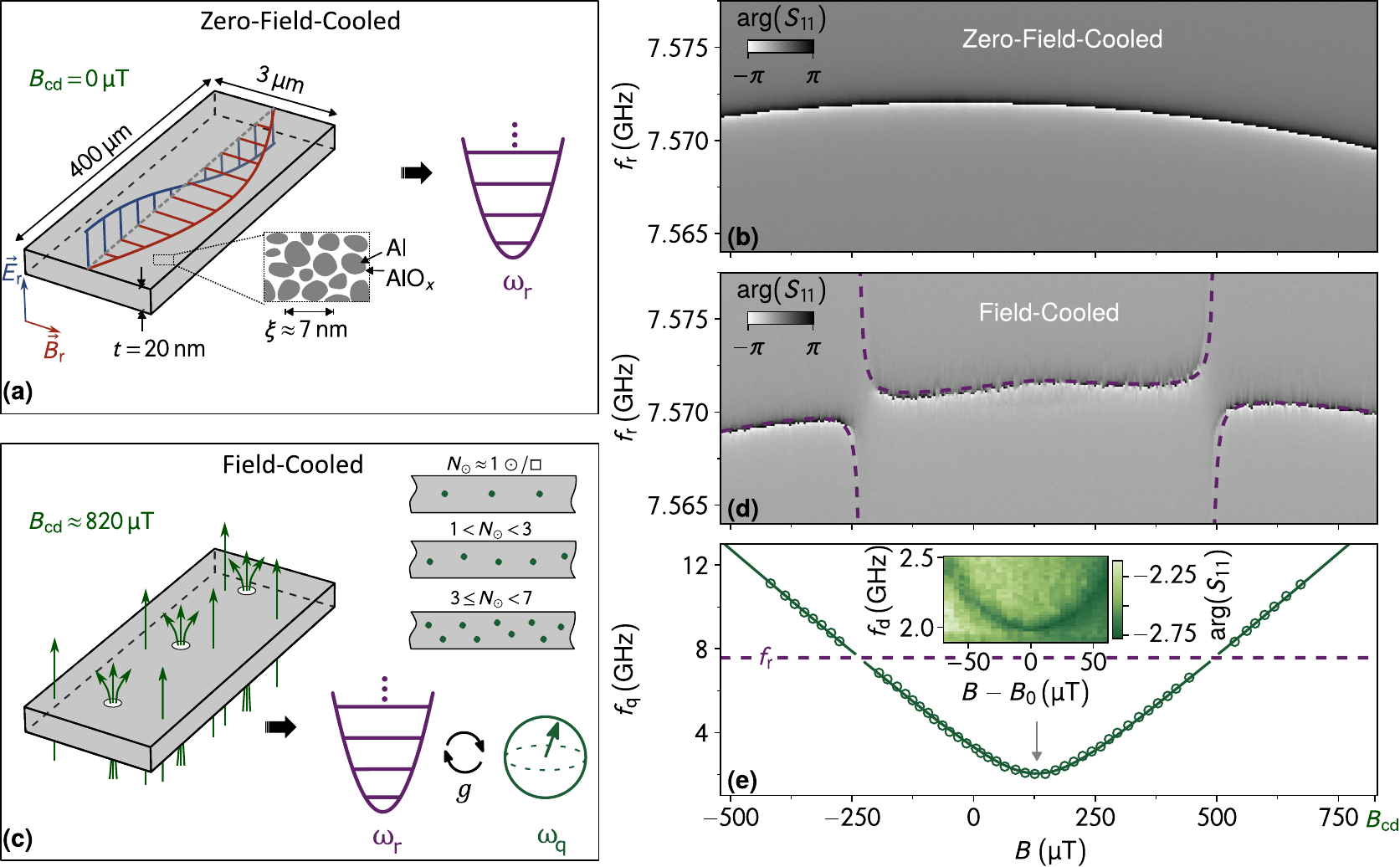}
    \caption{\label{fig: Fluxon}\textbf{Field cooling introduces vortex qubit states that couple to the grAl resonator.} 
    \textbf{(a)}~When cooled to $\qty{20}{\milli\kelvin}$ in perpendicular magnetic field $B_{\rm cd} = 0\,\si{\micro \tesla}$, a $\lambda/2$ microstripline grAl resonator behaves as a quantum harmonic oscillator. The electric and magnetic field distributions are illustrated in blue and red, respectively.
    \textbf{(b)}~Phase response $\arg(S_{11})$ of the resonator measured in reflection, as a function of perpendicular magnetic field $B$ applied after cooldown. The measured parabolic suppression of the resonance is given by the increase in kinetic inductance due to screening currents~\cite{annunziata2010tunable}, and the field range is limited by the vortex penetration threshold~\cite{borisov2020superconducting}.
    \textbf{(c)}~When cooled down in perpendicular magnetic field $B_{\rm cd} = \qty{820}{\micro \tesla}$, vortices enter the grAl resonator and the system exhibits a behavior akin to a flux qubit coupled to a readout resonator, as illustrated in panels (d) and (e). As shown in the top right corner, the number of vortices per square ($N_\odot$) determines their spatial arrangement. 
    \textbf{(d)}~The measured phase response of the resonator as a function of $B$ reveals avoided level crossings, suggesting coupling to vortex states. The purple dashed line shows a fit to the asymmetric quantum Rabi model (\cref{eq: Hamiltonian}), yielding the coupling $g/2\pi = \qty{92.5}{\mega\hertz}$. 
    \textbf{(e)}~Extracted VQ frequency $f_\text{q}$ from two-tone spectroscopy (see inset) as a function of $B$. The green line corresponds to the joint fit of data in (d) and (e) to \cref{eq: Hamiltonian}, and the purple dashed line marks the bare resonator frequency $f_r$. \textbf{Inset:} Two-tone spectroscopy in the vicinity of $B_{0}$ corresponding to the minimum frequency of the VQ. The colorscale indicates the measured phase response as a function of the frequency $f_d$ of the second drive.}
\end{figure*}

As schematized in \crefadd{fig: Fluxon}, we use a grAl micro-stripline resonator, with resistivity $\rho = \qty{3600}{\micro\ohm\centi\meter}$, chosen to be within a factor of three below the superconducting-to-insulating transition~\cite{Levy_2019}. 
In this regime, the film consists of Al grains of $\qtyrange{3}{4}{\nano \meter}$ diameter separated by amorphous $\mathrm{AlO_x}$ barriers, resulting in a coherence length $\xi\approx 7\,\si{\nano\meter}$ and London penetration depth of $\lambda_{\rm L} \approx \qty{4}{\micro \meter}$~\cite{Cohen_Abeles_1968, Deutscher_1973, Deutscher_2020}. 
The resonator is placed in a cylindrical copper waveguide (cf.~\cref{app:setup}) anchored to the $\qty{20}{\milli \kelvin}$ base plate of a dilution cryostat and measured in reflection. 
When cooled in zero magnetic field $B_{\mathrm{cd}}=0\,\si{\micro \tesla}$, the grAl resonator behaves as a weakly anharmonic oscillator~\cite{maleeva2018circuit}, with a fundamental frequency $\qty{7.572}{\giga \hertz}$, set by its dimensions ($\qty{3}{\micro \meter}$ wide, $\qty{400}{\micro \meter}$ long; cf.~\cref{app:sample}). 
The single-photon internal quality factor ($Q_i$) is $6\cdot 10^4$, similar to Ref.~\cite{Gru__Loss_Mech_grAl__2018}. 
\Crefadd{fig: Fluxon}{b} shows the frequency decrease with perpendicular magnetic field $B$, as expected with the increase in kinetic inductance~\cite{annunziata2010tunable,Eom2012wideband,borisov2020superconducting}. 

We introduce vortices into the grAl resonator through field cooling. 
Their formation and spatial arrangement depend on the value of the flux bias during cooldown $\phi = B_{\rm cd} w^2/\Phi_0$, where $\Phi_0=h/2e$ is the magnetic flux quantum and $w$ is the width of the resonator. 
In the Pearl limit~\cite{Pearl_1964}, where the thickness of the film $t \ll \lambda_L$, the threshold for stable vortices
is $\phi_{\rm S} = 2/\pi\ln{(2w/\pi\xi)}$~\cite{Kogan_1994, stan2004critical, Bronson_Gelfand_Field_2006}, corresponding to $\phi_{\rm S}=3.57$ for our geometry. In the range $\phi_{\rm S} <\phi < 2.48 \, \phi_{\rm S}$, each square of the film is threaded on average by 1 to 3 vortices ($1\!\leq\!N_\odot\!<\!3$).  
For $\phi > 2.48\,\phi_{\rm S}$~\cite{Bronson_Gelfand_Field_2006}, the vortex state undergoes the first buckling transition in which the vortex chain splits into a two-row configuration, as sketched in \crefadd{fig: Fluxon}{c} (upper inset). 
To avoid the complex vortex dynamics above the buckling transition, all measurements presented in this manuscript are carried out after field cooling in a perpendicular field of $B_{\rm cd} = \qty{820}{\micro\tesla}$, corresponding to $\phi = 1.0\,\phi_s$ (or $ 1\!\leq\!N_\odot\!<\!3$).

\begin{figure*}[htbp]
    \includegraphics[width=1.0\textwidth]{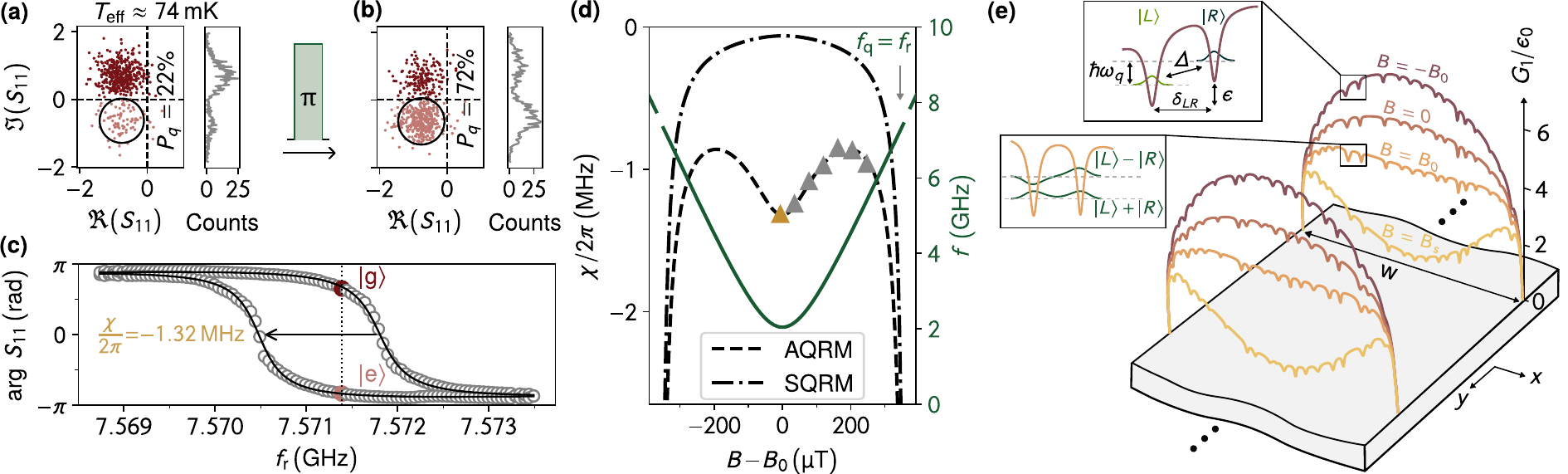}
    \caption{\label{fig: TD_Chi}\textbf{The asymmetric quantum Rabi model describes the vortex qubit dispersively coupled to its resonator.} 
    \textbf{(a)}~Consecutive $S_{11}$ measurements at the sweet spot show two IQ-clouds in the complex plane. The relative occurrence of points in the clouds corresponds to the population of the $\ket{g}$ (ground) and $\ket{e}$ (excited) states and yields an effective qubit temperature $T_{\rm eff} \approx \qty{74}{\milli\kelvin}$. 
    \textbf{(b)}~Measured IQ-clouds following a $\qty{20}{\nano\second}$ drive at $f_q$ calibrated to implement a $\pi$-pulse, show a population inversion as expected for a two-level system. The black circles have a radius of $1.5$ standard deviation. 
    \textbf{(c)}~Resonator phase response arg($S_{11}$), obtained from the centers of the IQ-clouds, measured versus readout frequency in the vicinity of $f_r$. A fit to the data (black solid line) yields a dispersive shift of $\chi/2\pi = \qty{-1.32}{\mega\hertz}$. The dark red ($\ket{g}$) and light red ($\ket{e}$) points correspond to the data in (a) at $f_{\rm{RO}} = \qty{7.5714}{\giga\hertz}$ (dashed line).  
    \textbf{(d)}~Variation of $\chi$ with magnetic field $B$, shown as triangles, with the yellow triangle corresponding to the measurement in (b). The dashed line indicates the expected values from the asymmetric quantum Rabi model \cref{eq: Hamiltonian} with $g_{\rm AQRM}/2 \pi = \qty{92.5}{\mega \hertz}$, and the dash-dotted line to the symmetric quantum Rabi model \cref{eq: SQRM} with $g_{\rm SQRM} /2 \pi = \qty{20}{\mega \hertz}$. The solid green line represents the qubit frequency (right axis), similar to \crefadd{fig: Fluxon}{d}.
    \textbf{(e)}~Gibbs free energy $G_1$ (cf.~\cref{eq:Gibbs}, baseline) of a single vortex along the width of the resonator in units of $\varepsilon_0 = \Phi_0^2/2\pi\mu_0\Lambda \approx 2\,\si{\tera \hertz}$, with added pinning potentials depicted as Lorentzian dips. The colors correspond to different applied magnetic fields from $B_{\rm S} = \phi_{\rm S}\Phi_0/w^2$ to $-B_0$.
    \textbf{Top inset:}~A possible double-well potential arising from the energy landscape of neighboring pinning sites $\delta_{\rm LR}$ apart, offset in energy by $\epsilon$. The localized wavefunctions in these wells correspond to the two possible positions $\ket{\rm L}$ and $\ket{\rm R}$ of the vortex, coupled by tunneling amplitude $\Delta$, with an energy separation given by the VQ transition frequency $\hbar\omega_\mathrm{q}$. 
    \textbf{Bottom inset:}~the double-well potential is degenerate at the sweet spot, with VQ states depicted as symmetric and antisymmetric combinations of the localized wavefunctions, and $\hbar\omega_\mathrm{q}=2\Delta$.
    }
\end{figure*}

Following field cooling, sweeping $B$ reveals avoided level crossings in the grAl resonator response as illustrated in \crefadd{fig: Fluxon}{d}, which we interpret as evidence of strong coupling with $g/2\pi = \qty{92.5}{\mega\hertz}$ to vortex states. 
To extract the mode’s spectrum, we sweep a second microwave drive while probing the readout resonator (see \crefadd{fig: Fluxon}{e}). 
We observe a minimum vortex mode frequency $f_{\rm q} = 2\,\si{\giga\hertz}$ at the sweet spot $B_0 = \qty{128}{\micro\tesla}$ (cf.~\crefadd{fig: Fluxon}{e} inset), with a hyperbolic field dispersion eccentricity $\gamma = \qty{20}{\giga \hertz / \milli \tesla}$, reminiscent of a flux qubit~\cite{Chiorescu_2003}.
As the field approaches the sweet spot, the resonance narrows, pointing to magnetic field fluctuations as dominant noise source~\cite{Faoro2008Jun}.
Repeated field coolings typically result in a single prominent avoided crossing, though we occasionally observe zero or multiple.
From measured spectra across 32 field-cooling cycles in six different resonators, we extract values of $g$, $f_{\rm q}$, $B_0$, and $\gamma$ that are of similar order of magnitude but vary between cycles (cf.~\cref{app:Multiple_traps}), suggesting different underlying vortex configurations.
\newpage
Repeated resonator $S_{11}$ measurements at the sweet spot reveal two distinct clusters in the quadrature plane (\crefadd{fig: TD_Chi}{a}), indicating that the vortex state has a lifetime well beyond the $\qty{1.2}{\micro \second}$ integration time, thereby enabling single-shot state discrimination.
As demonstrated in \crefadd{fig: TD_Chi}{b}, by driving at $f_{\rm q}$ we can calibrate a $20\,\si{\nano\second}$ $\pi$-pulse, which inverts its thermal population (see~\cref{app:Rabi} for the Rabi oscillations). 
These signatures define the VQ states $\ket{g}$ and $\ket{e}$. 
From their steady-state populations we extract a $\qty{74}{\milli\kelvin}$ effective temperature.
The VQ-resonator interaction induces a state-dependent dispersive shift $ \chi/2\pi = f_{{\rm r}, \ket{e}} - f_{{\rm r}, \ket{g}} $. 
As shown in \crefadd{fig: TD_Chi}{c}, fitting the resonator's phase response to the centers of IQ clouds measured versus readout frequency yields $\chi/2\pi = \qty{-1.32}{\mega\hertz}$ (see~\cref{app:Dispersive measurements} for all measured IQ clouds).

\begin{figure*}[t!]
    \includegraphics[width=1\textwidth]{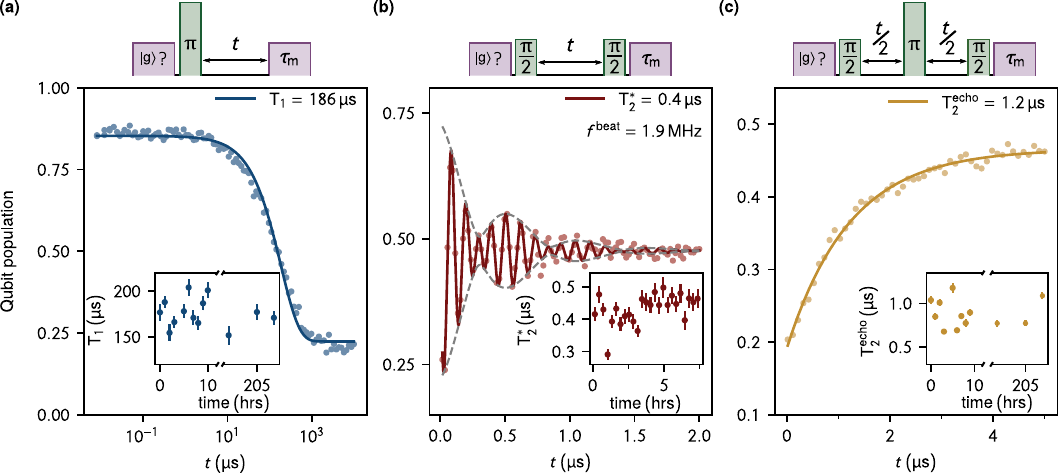}
    \caption{\label{fig: Times}\textbf{Measurement of low loss and coherence in the VQ.} 
    \textbf{(a)}~Free energy decay measured after a $20\,\si{\nano \second}\,\,\pi$ pulse applied selectively to the VQ measured in the ground state $\ket{g}$. The readout pulse has a duration $\tau_\text{m}=1.2 \,\si{\micro \second}$. The excited vortex qubit population P($\ket{e}$) as a function of wait time $t$ is fitted with an exponential corresponding to $T_1 = 186\,\si{\micro \second}$ (solid line). 
    \textbf{(b)}~Ramsey fringes exhibit a beating pattern, resulting from two frequencies separated by $f^{\rm beat} = 2\,\si{\mega \hertz}$. We extract $T_2^*$ Ramsey coherence times of $440\,\si{\nano \second}$. 
    \textbf{(c)}~Spin Hahn echo measurement with extracted $T_{2}^{\text{echo}} = 1.2\,\si{\micro\second}$. For each panel, the corresponding pulse sequence is sketched at the top, and the insets show measured coherence times over several hours.}
\end{figure*}

For further insight into the nature of the VQ and its interaction with the grAl resonator, we measure $\chi$ versus field, as shown in \crefadd{fig: TD_Chi}{d}. We model it using the quantum Rabi model (QRM) for a spin S = $1/2$ coupled via $\mathcal{H}_{\rm c} = \hbar g (\hat{a}^{\dagger} + \hat{a})\sigma_{x}$ to a harmonic oscillator with frequency $\omega_\text{r}$ and Hamiltonian $\mathcal{H}_\text{r} = \hbar \omega_\text{r} \left(\hat{a}^{\dagger} \hat{a} + \tfrac{1}{2} \right)$ (see \cref{app:model}). 
Here $\hat{a}^{\dagger}$ and $\hat{a}$ are the resonator bosonic operators and $\sigma_x$ is the Pauli matrix for a spin $\bm{S}=\hbar/2\,\bm{\sigma}$.
The interaction energy between the spin and the magnetic field is \mbox{$\gamma \bm{S}\cdot(\tilde{\bm{B}} + \bm{B}')$}, where $\gamma$ is the gyromagnetic ratio and the field consists of two contributions: a pseudo-field $\tilde{\bm{B}}$ that sets the VQ energy at the sweet spot, and the applied magnetic field $|\bm{B}'| = B - B_{0}$ measured from the sweet spot.
We compare joint fits of the measured VQ and resonator frequencies in field (cf.~\crefadd{fig: Fluxon}{d,e}), using the symmetric
\begin{equation}
    \mathcal{H}_{\rm SQRM}  = \mathcal{H}_\text{r} + \mathcal{H}_\text{c} +  \frac{\hbar\gamma}{2}\sigma_z \cdot \sqrt{\tilde{B}^2 + {B'}^2}\,, \label{eq: SQRM}
\end{equation}
and the asymmetric QRM
\begin{align}
    \mathcal{H}_{\rm AQRM}  &= \mathcal{H}_\text{r} + \mathcal{H}_\text{c} + \frac{\hbar\gamma}{2}\sigma_z \cdot \tilde{B} -  \frac{\hbar\gamma}{2}\sigma_x \cdot B' \label{eq: Hamiltonian}\,.
\end{align}
Only the AQRM captures the non-monotonic dependence of $\chi$ with $B$. In contrast, the SQRM predicts a monotonically decreasing $\chi$ with detuning from the resonator.  
Moreover, using the coupling constant $g$ from the joint fit in \crefadd{fig: Fluxon}{d,e}, we obtain quantitative agreement for the measured $\chi$, as demonstrated in \crefadd{fig: TD_Chi}{d}.  
This suggests that the VQ, possibly consisting of persistent currents, arises from dynamics in a double-well potential, analogous to fluxon tunneling through the Josephson junction of a flux qubit~\cite{Chiorescu_2003}. Within this model, the pseudo field $\tilde{B}$ is given by the fluxon tunneling amplitude~\cite{Matveev2002}.

To give a hypothesis for the origin of the double-well potential, consider the vortex Gibbs energy~\cite{Kuit_2008} in our geometry (cf.~\cref{fig: Fluxon}):
\begin{equation}
    G_1(x) = \varepsilon_0 \ln{\left(\frac{2w}{\pi \xi} \sin\left(\frac{\pi x}{w}\right) + 1\right)} - \frac{\Phi_0 (B - n\Phi_0)}{\mu_0 \Lambda}x(w-x)\,,
    \label{eq:Gibbs}
\end{equation}
where $\varepsilon_0 = \Phi_0^2/(2\pi\mu_0\Lambda)$ sets the single-vortex energy scale, $n$ is the density of vortices ($n = 0$ for the first vortex), $\Lambda = 2\lambda_{\rm L}^2/t$ is the Pearl length of the resonator, and $x$ is the position of the vortex.
As $B$ decreases from $B_{\rm S} = \phi_{\rm S}\Phi_0 /w^2$ to zero, the minimum of $G_1(x)$ vanishes (cf.~\crefadd{fig: TD_Chi}{e}, baseline), and in the absence of pinning the vortex would be expelled.
To account for the measured stability of the VQ across magnetic field sweeps (cf.~\cref{fig: Fluxon}), we incorporate pinning potentials, presumably abundant given the disordered nature of grAl.
They are modeled by adding Lorentzian dips $V_{\rm pin} = V_i \left(1 +(x-x_i)^2/\sigma_i^2\right)^{-1}$ to $G_1(x)$, at random positions $x_i$, depth $V_i$, and width $\sigma_i$, sketched as the colored energy landscapes in \crefadd{fig: TD_Chi}{e}.
A vortex tunneling between pinning sites forms a double-well potential (cf.~\crefadd{fig: TD_Chi}{e} top inset), in which $B$ tunes the relative pinning depths according to~\cref{eq:Gibbs}.
At $B_0$ the minima are degenerate and the vortex delocalizes, with $\ket{g}$ and $\ket{e}$ given by symmetric and antisymmetric superpositions of $\ket{L}$ and $\ket{R}$ wavefunctions (cf.~\crefadd{fig: TD_Chi}{e} bottom inset). 

This hypothesis is supported by the fact that typically measured gyromagnetic ratios $\gamma/2\pi=\qtyrange{3}{25}{\giga\hertz/\milli\tesla}$ are consistent with flux tunneling between pinning sites separated by tens of nanometers (cf.~\cref{app:hyperbolic_freq}), reminiscent of tunneling through grAl nanojunctions~\cite{Rieger_Gralmonium_2022}.
Moreover, to leading order, a kinetic-inductance-mediated VQ-resonator coupling $g/\omega_{\rm r} \simeq \qtyrange{0.1}{1}{\percent}$~(cf.~\cref{app:understanding}) is consistent with the observed avoided level crossings.
While single-vortex pinning can account for the observed VQ, it is well established that multiple vortices simultaneously enter the resonator once the threshold for entry is reached~\cite{Bronson_Gelfand_Field_2006}, as illustrated by the set of Gibbs curves in the foreground of \crefadd{fig: TD_Chi}{e}. 
We estimate the VQ-VQ interaction in the \qtyrange{10}{100}{\mega \hertz} range (cf.~\cref{app:vortex-vortex}), suggesting collective vortex dynamics as unlikely.
Nevertheless, distinguishing between single- and multi-vortex dynamics, for instance using imaging methods~\cite{Kirtley2010Nov, Suderow_2014, Persky2022Mar, Weber2025Aug, Bai_2025}, or by shaping the resonator width~\cite{nsanzineza2014trapping} remains an important avenue for future research.

We complete the characterization of the vortex qubit with time-domain measurements at the sweet spot. 
As shown in~\crefadd{fig: Times}{a}, the fitted energy relaxation time is $T_1=\qty{186}{\micro\second}$, with values ranging from $\qtyrange{40}{300}{\micro \second}$ across multiple VQ preparation cycles (cf.~\cref{app:more_info}).
Relaxation times extracted from VQ quantum jumps (cf.~\cref{app:more_info}) fall within the temporal fluctuations observed in free decay, indicating a quantum non-demolition readout.
Remarkably, the VQ exhibits quantum coherence, with a Ramsey time $T_{2}^{*}=\qty{440}{\nano\second}$, which extends to $T_{2}^{\rm echo}=\qty{1.2}{\micro\second}$ in Hahn-echo measurements which suppress the low-frequency noise (cf.~\crefadd{fig: Times}{b,c}).
The Ramsey fringes exhibit a beating pattern, corresponding to a toggling of the VQS frequency between two values separated by $\qty{1.9}{\mega\hertz}$. This feature is sometimes also observed in superconducting qubits~\cite{Rieger_Gralmonium_2022, stern2014flux}, possibly indicative of charge noise or conductance channel fluctuations.
In future experiments, detailed noise characterization~\cite{Yan_2016}, environment polarizability~\cite{Spiecker_2023}, in-plane magnetic\cite{Günzler_2025} and electric field~\cite{Sueur2018Oct, Kristen_2024} susceptibility, etc., could shed light on the microscopic origin of the VQ and its environment.

In conclusion, field-cooling a granular aluminum micro-stripline resonator reproducibly generates vortex qubit states that couple dispersively to the resonator and can be coherently driven. 
Our results demonstrate that superconducting vortices can harbor quantum coherence on microsecond timescales.
Remarkably, the VQ energy relaxation times are on the order of hundreds of microseconds, comparable to those of engineered superconducting qubits~\cite{Kjaergaard_Oliver_2020, Blais__Review_cQED__2021}, and qualitatively distinct from the dissipation expected for Abrikosov vortex dynamics.
This supports a picture of grAl as a 3D network of Josephson junctions, capable of hosting gapful-vortices~\cite{kiselov2023gapful}.
The observed dispersive shifts and spectra are accurately captured by an asymmetric quantum Rabi model, consistent with a two-level system in a double-well potential.
Microscopically, this may arise from vortex tunneling between pinning sites, modulated by the magnetic field dependence of the Gibbs energy.
This hypothesis, while consistent with our measurements, remains to be confirmed by future experiments such as scanning tunneling or scanning SQUID microscopy. 

Looking ahead, the measurement of quantum coherence in vortex states, along with their relative technological simplicity, opens several exciting avenues in quantum science.
Disordered superconductors beyond granular aluminum~\cite{Weitzel_2023, Charpentier_2025} or engineered 2D networks of Josephson junctions~\cite{Bottcher2023Oct, Bottcher2024Nov} may host similar VQs, shedding new light onto the complex physics in the vicinity of the superconductor to insulator transition~\cite{Haviland1989May, GlezerMoshe2021Aug}.
Moreover, this would offer an embedded tool for material characterization at the microscopic level.
In the same spirit, if the observed dynamics indeed stem from single-vortex tunneling, VQs could be harnessed for nanoscale sensing.
Ultimately, engineering the pinning landscape and device geometry, combined with noise spectroscopy and susceptibility measurements to magnetic and electric fields, will be crucial to enhance VQ coherence and possibly launch a vortex-based quantum information platform.

\section*{Data Availability}
All relevant data are available from the corresponding author upon reasonable request.
\section*{Acknowledgements}
The authors are grateful to M. Feigel'man, L. Glazman, G. Kirchmair, N. Roch, A. Shnirman, U. Vool, and W. Wulfhekel for fruitful discussions and acknowledge the technical support from L. Radtke and S. Diewald. The authors acknowledge funding from the Baden-Württemberg Stiftung under the QT-10 project (QEDHiNet). The KIT Nanostructure Service Laboratory provided support for the facilities used. We recognize the qKit measurement software framework. The authors acknowledge support by the state of Baden-Württemberg through the bwHPC. M.S. and N.G. acknowledge support from the German Ministry of Education and Research (BMBFTR) within the GEQCOS project (FKZ: 13N15683). N.Z. acknowledges funding from the Deutsche Forschungsgemeinschaft (DFG – German Research Foundation) under project number 450396347 (GeHoldeQED). S.Gü., D.R., and W.W. acknowledge support from the Leibniz award WE 4458-5.

\bibliography{Quantum_vortices}
\balancecolsandclearpage

\onecolumngrid
\section*{Appendices}
\vspace{0.6cm}
\twocolumngrid
\appendix
\section{\label{app:setup}Setup \& Magnetic Field calibration}
\begin{figure}[tb]
    \includegraphics[width=1.0\columnwidth]{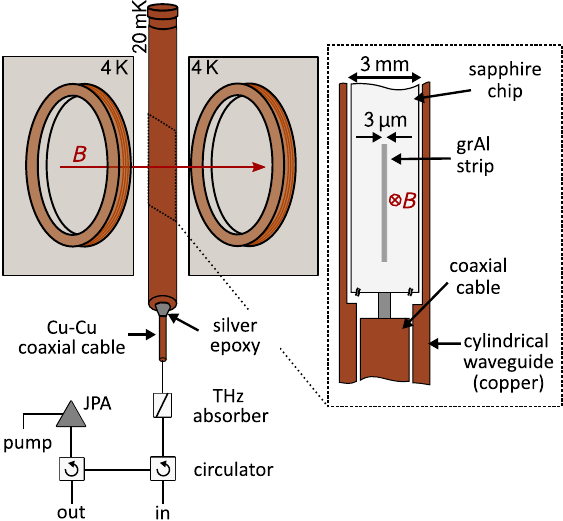}
    \caption{\label{fig: setup} \textbf{Measurement setup.} 
    Schematic of the cylindrical copper waveguide sample holder anchored to the millikelvin stage.
    A Helmholtz coil pair, thermalized to the \qty{4}{\kelvin} stage of the cryostat, generates a perpendicular magnetic field $B$.
    The sample holder, directly connected to a \unit{\tera\hertz} absorber~\cite{Rehammar_Gasparinetti_2023}, is measured in single-port microwave reflection $S_{11}$, with a Josephson parametric amplifier (JPA)~\cite{Winkel__Amplifier__2020} on the output line.
    \textbf{Inset:}~A $3 \times 10 \, \si{\milli \meter}$ sapphire chip is secured with a copper dowel and hosts a grAl micro-stripline resonator (gray rectangle) positioned $\qty{0.5}{\milli \meter}$ from the bottom edge of the chip, similar to  Refs.~\cite{borisov2020superconducting, Günzler_2025}.
    }
\end{figure}

\begin{figure}[tb]
    \includegraphics[width=1.0\columnwidth]{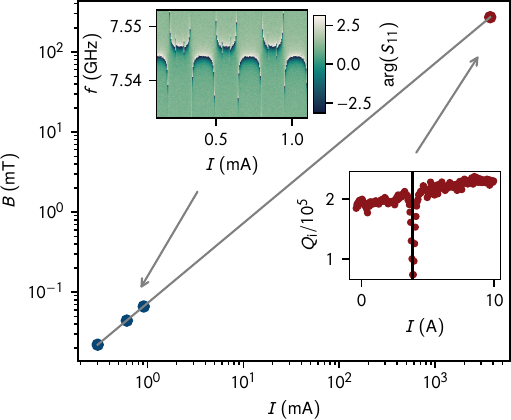}
    \caption{\label{fig: coil} \textbf{Magnetic field calibration.}
    Extracted field $B$ at the sample position versus bias current applied to the Helmholtz coil.
    For fields up to $\qty{100}{\micro \tesla}$, the flux periodicity (flux sensitivity of $\qty{22}{\micro \tesla/\Phi_0}$) of a Gralmonium flux qubit~\cite{Rieger_Gralmonium_2022} is used for the calibration. 
    The blue markers denote fields corresponding to integer multiples of $\Phi_0$ in the qubit loop.
    At higher fields, calibration relies on electron spin resonance (ESR) of $g=2$ spin-$1/2$ paramagnetic impurities (red marker).
    A linear fit to both data sets yields a conversion factor of $\qty{72.8}{\milli \tesla}/\unit{\ampere}$.
    \textbf{Top inset:}~Phase response $\arg(S_{11})$ of the readout resonator plotted as a function of the drive frequency $f$ at different bias currents $I$ for the gralmonium qubit described in Ref.~\cite{rieger2023fano}. 
    The avoided level crossings appear at field values where the qubit mode periodically crosses the resonator frequency.
    \textbf{Bottom inset:}~Internal quality factor $Q_{\rm i}$ of a grAl resonator versus current $I$ in the coil. 
    A dip, marked by the black line, appears at the current corresponding to ESR between the $g=2$ spins and the resonator. 
    }
\end{figure}

\Cref{fig: setup} shows the cylindrical sample holder with a \qty{3}{\milli \meter} diameter and a \qty{60}{\giga \hertz} cut-off frequency. 
The resonator is coupled to the microwave readout and control electronics via the evanescent field of the coaxial pin, as illustrated in the inset of \cref{fig: setup}.
We calibrate the magnetic field at the sample position using two complementary methods (cf.~\cref{fig: coil}). 
At low fields, $B<\qty{100}{\micro\tesla}$, we exploit the flux periodicity of a Gralmonium flux qubit, set by the qubit loop area of $\qty{94}{\micro \meter}^2$. 
The qubit is identical in design to Ref.~\cite{Rieger_Gralmonium_2022} and is mounted in the same sample holder, instead of the grAl resonators discussed in the main text.
At fields $B>\qty{100}{\milli\tesla}$, calibration is performed using electron spin resonance (ESR) of $g = 2$ spin$-1/2$ paramagnetic impurities in grAl resonators, commonly observed in the environment of superconducting devices~\cite{borisov2020superconducting, Günzler_2025}. 
When the spins are tuned into resonance with the grAl resonator ($f_{\rm r} = \qty{7.627}{\giga \hertz}$), we observe enhanced microwave losses, manifesting as a dip in the internal quality factor $Q_{\rm i}$ at a magnetic field $B = hf_{\rm r}/g\mu_\mathrm{B} = \qty{272.5}{\milli \tesla}$ (cf.~\cref{fig: coil} bottom inset).
Following Ref.~\cite{borisov2020superconducting}, to minimize vortex trapping and resonator frequency shift during this sweep, the chip is aligned parallel to the magnetic field.

\section{\label{app:sample}Sample \& Fabrication Details}
\begin{figure*}[t!]
    \includegraphics[width=1\textwidth]{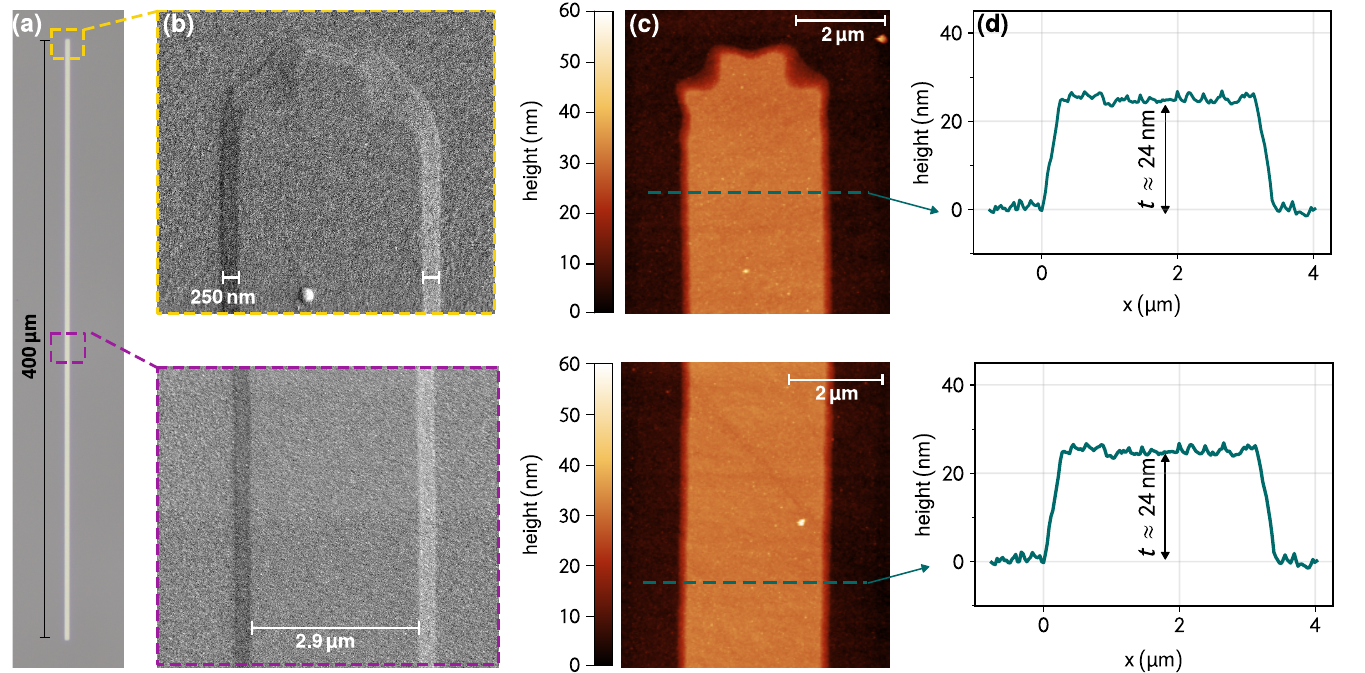}
    \caption{\label{fig: Image} 
    \textbf{Microscopy of the grAl resonator.}
    \textbf{(a)}~Optical micrograph showing the $\qty{400}{\micro \meter}$ long resonator.
    \textbf{(b)}~Scanning electron microscope (SEM) image of the resonator’s end (top) and center (bottom) reveals $\qty{250}{\nano \meter}$ wide slanted edges.
    \textbf{(c)}~Atomic force microscopy (AFM) image acquired in tapping mode (tip radius $< \qty{8}{\nano \meter}$).
    \textbf{(d)} The AFM line profile yields a film thickness of $t = 24\,\si{\nano \meter}$.}
\end{figure*}
We fabricate the sample on a double-side polished c-plane sapphire substrate using electron-beam lithography and wet etching.
The substrate is cleaned in a $\qty{50}{\celsius}$ acetone bath and rinsed in ethanol, before applying an $\mathrm{Ar/O_{2}}$ ion descum process using a Kaufman ion source in a PreVac evaporation system.
After performing titanium gettering, we deposit a \qty{20}{\nano\meter} grAl film at room temperature by evaporating aluminum at \qty{1}{\nano\meter/\second} under dynamic oxidation.
The resulting film has a sheet resistance of $\qty{1.5}{\kilo \ohm / \Box}$.
To define the resonator geometry, we pattern a  $\qty{300}{\nano \meter}$ thick resist layer (ARN 7520.18) using a $50\,\si{\kilo \electronvolt}$ e-beam writer.
We develop the resist for $\qty{40}{\second}$ in an AR 300-47:H$_2$O mixture (4:1), followed by wet etching in MF 319.
\Cref{fig: Image} shows microscopy of the resulting grAl resonator structures.
The resonator has a uniform height of \qty{24}{\nano \meter} with slanted edges.

\vspace{0.3cm}

\section{\label{app:Multiple_traps} Repeated initialization of VQs}
\begin{figure*}[tb]
    \includegraphics[width=1.0\textwidth]{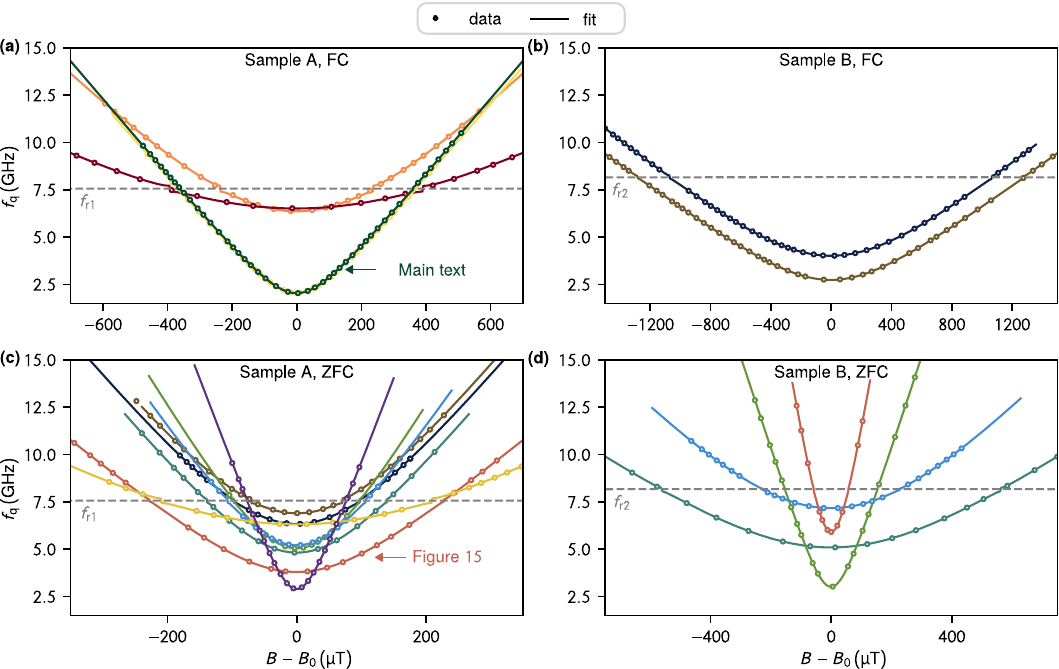}
    \caption{\label{fig: Alltraps}
    \textbf{Spectroscopy of additional VQs.}
    Extracted VQ frequency with fits to the Hamiltonian (\cref{eq: Hamiltonian}) as a function of magnetic field detuning from the sweet spot $B_0$.
    Different colors indicate independent preparation cycles.
    Data are shown for two resonators fabricated from the same grAl film and identical geometry: Sample A (\textbf{a,c}, $f_{\rm r1} = \qty{7.5}{\giga\hertz}$) and Sample B (\textbf{b,d}, $f_{\rm r2} = \qty{8.1}{\giga\hertz}$).
    VQs are introduced by field cooling (top panels) or by zero-field cooling followed by ramping to 1 mT (bottom panels). Gray dashed lines mark the corresponding bare resonator frequencies.
    }
\end{figure*}

\begin{figure}[tb]
    \includegraphics[width = 1.0\columnwidth]{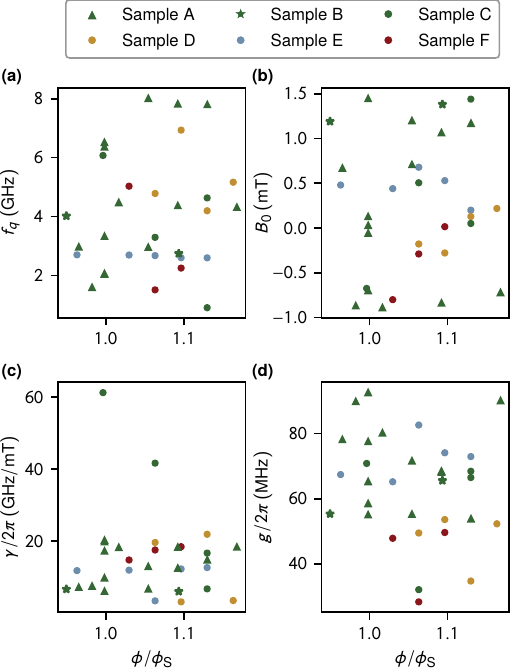}
    \caption{\label{fig: correlations}\textbf{Dependence of VQ parameters on bias field.}
    \textbf{(a)}~Sweet-spot qubit frequency $f_\mathrm{q}$, \textbf{(b)}~offset field $B_0$, \textbf{(c)}~gyromagnetic ratio $\gamma$, and \textbf{(d)} VQ-resonator coupling $g$ as a function of the field bias $\phi$, normalized to $\phi_{\rm S}$ (cf.~main text and Ref.~\cite{Kogan_1994}).
    Parameters are extracted from fits to \cref{eq: Hamiltonian}.
    Spectroscopy for Samples A and B is shown in \cref{fig: Alltraps}. 
    Samples C–F were fabricated from a grAl film deposited in a different e-beam evaporator compared to the rest. 
    Samples A–C have the same dimensions (see \cref{fig: Image}), while the resonator length increases from $\qty{400}{\micro\meter}$ (Sample C) to $\qty{720}{\micro\meter}$ (Sample F).
    }
\end{figure}

\Crefadd{fig: Alltraps}{a} shows VQ field spectra from four field-cooling (FC) preparation cycles, each involving thermal cycling of the grAl resonator to $T> \qty{10}{\kelvin}>T_\mathrm{c}$ followed by field cooling at $B_{\rm cd} = \qty{820}{\micro \tesla}$.
We reproducibly observe hyperbolic VQ spectra with varying sweet spot frequency and eccentricities.
Similar results are obtained by field cooling other grAl resonator samples (cf.~\crefadd{fig: Alltraps}{b}).
Interestingly, VQs can also be introduced by zero-field-cooling (ZFC) the grAl resonator, followed by applying a field $B\approx \qty{1}{\milli \tesla}$ at base temperature.
As shown in \crefadd{fig: Alltraps}{c} and \crefadd{fig: Alltraps}{d}, the resulting spectra for VQs induced in ZFC samples exhibit hyperbolic characteristics comparable to those from FC.
Previous observations of vortex trapping in ZFC grAl resonators (Ref.~\cite{borisov2020superconducting}) can now be suspected to also give rise to VQs.

Including the sample characterized in the main text, we measured a total of 32 VQ-resonator spectra prepared via field cooling in six grAl resonators across two wafers. 
In \cref{fig: correlations}, we show the extracted VQ-resonator characteristics and find no correlation with the cool-down field.

\section{\label{app:Rabi} Rabi Oscillation and Active State Preparation}
\begin{figure}[tb]
    \includegraphics[width=1.0\columnwidth]{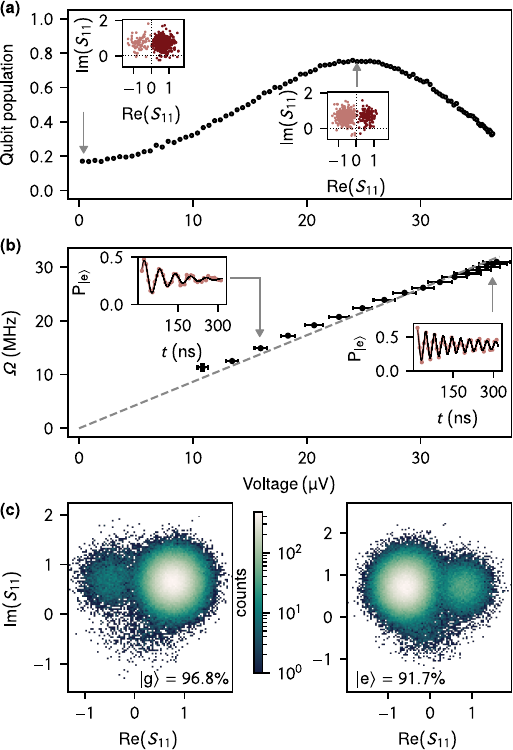}
    \caption{\label{fig:rabi} \textbf{Coherent control of the VQ.}
    \textbf{(a)}~Rabi oscillations of the excited-state population (black circles) as a function of the drive amplitude at the sample holder. 
    \textbf{Insets:}~Scatter plots of repeated $S_{11}$ measurements in the complex plane, normalized to the mean $|S_{11}(\ket{g})|$, show population inversion of the ground (dark red) and excited (light red) state population between drive amplitudes of \qty{0}{\micro \volt} and \qty{25}{\micro \volt} (left and right inset, respectively).
    \textbf{(b)}~Extracted Rabi frequency $\Omega$ (black circles) versus drive amplitude with a linear fit (gray line). Error bars indicate the uncertainty in the power delivered to the sample holder. 
    \textbf{Insets:}~Rabi oscillations of the excited-state population  $\rm{P_{\ket{e}}}$ as a function of pulse duration $t$ for rectangular pulses with amplitudes of $\qty{16}{\micro\volt}$ (left inset) and $\qty{36.3}{\micro\volt}$ (right inset). 
    Black lines indicate fits to a damped cosine function.
    \textbf{(c)}~IQ histograms of \num{2e5} single-shot active state preparations in the ground (left) and excited (right) states, using conditional $\pi$-pulses triggered on detection of $\ket{e}$ or $\ket{g}$, respectively.
    }
\end{figure}

\Cref{fig:rabi} shows Rabi oscillations of the excited-state population, confirming coherent quantum TLS operation.
While thermal initialization yields a population inversion of $70\%$, 
conditional $\pi$-pulses enable ground- and excited-state preparation fidelities of \qty{96.8}{\percent} and \qty{91.7}{\percent}, limited by readout fidelity. 

\section{\label{app:Dispersive measurements} Dispersive measurements}
\begin{figure*}[tb]
    \includegraphics[width=1.0\textwidth]{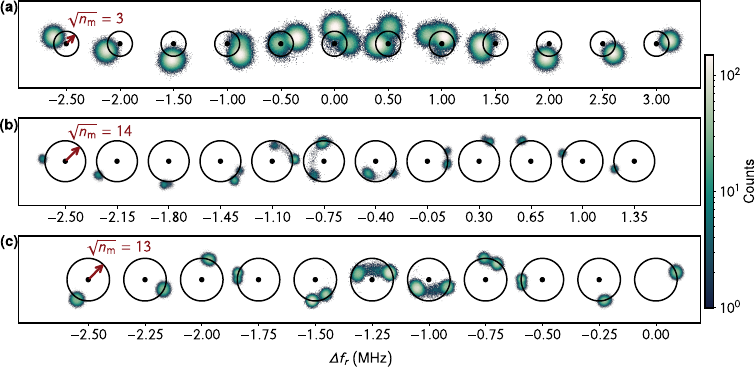}
    \caption{\label{fig: Dispersive} 
    \textbf{IQ histograms versus readout detuning.}
    IQ histograms, plotted as a function of readout detuning $\Delta f_\mathrm{r}$ from the resonator frequency.
    Black circles indicate the complex resonator response for a readout pulse corresponding to $n_\mathrm{m}$ measurement photons.
    Measurements are shown for \textbf{(a)}~the FC VQ in Sample A (main text VQ), \textbf{(b)}~the VQ induced in ZFC Sample A (cf.~\cref{fig: Alltraps}), and \textbf{(c)}~the FC VQ in Sample C (cf.~\cref{fig: correlations}).
    }
\end{figure*}

In \cref{fig: Dispersive} we show IQ histograms used to extract the dispersive shift $\chi$ (cf.~\cref{fig: TD_Chi,Fig: another_trap}). 
The two IQ clouds corresponding to $\ket{g}$ and $\ket{e}$ rotate along the $S_{11}$ circle in the IQ plane, with a radius set by the square root of the measurement photon number~\cite{Hatridge2013Jan, vool2014non}. 
The absence of additional clouds indicates a well-isolated computational qubit basis.

\clearpage
\section{\label{app:model} Asymmetric Quantum Rabi Model}

\begin{figure}[!h]
    \includegraphics[width = 0.95\columnwidth]{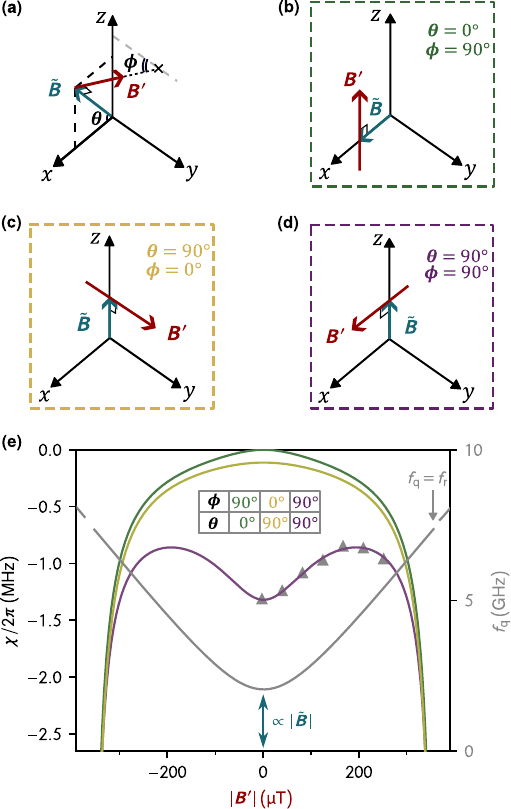}
    \caption{\label{fig:fitting} 
    \textbf{Field orientations in the AQRM Hamiltonian.}
    \textbf{(a)}~Coordinate system defining the orientation of the pseudo-field $\tilde{\bm B}$ and the applied field relative to the sweet spot $\bm{B}'$ for the $\sigma_x$ interaction in the AQRM (\cref{eq: general}). 
    The pseudo-field $\tilde{\bm B}$ lies in the $xz$-plane at angle $\theta$ to the $x$-axis, and is orthogonal to $\bm {B}'$, which is tilted at an angle $\phi$ to the $y$-axis. Note that $z$ and $x$ coordinates here do not coincide with the spatial directions of the field in the laboratory frame.
    \textbf{(b-d)}~Representative orientations of $\tilde{\bm B}$ and $\bm{B}'$ used to model the VQ-resonator system via the AQRM (\cref{eq: general}). 
    All other configurations can be mapped onto or constructed from these cases.
    \textbf{(d)}~Measured dispersive shift $\chi$ (triangle markers, left axis) and qubit frequency $f_\mathrm{q}$ (gray line, right axis) as a function of $\lvert \bm{B}' \rvert$.
    The predicted $\chi$ for the field parameterizations ($\phi$, $\theta$) of the AQRM in panels (b-d) are indicated as lines in green, yellow, and purple.
    Data for $\chi$ and $f_\mathrm{q}$ are identical to \crefadd{fig: TD_Chi}{d}.
    }
\end{figure}

In the following, we justify the specific form of the AQRM in \cref{eq: Hamiltonian} used to model the qubit-resonator system.
To reproduce the observed hyperbolic qubit spectrum, two orthogonal field components are required: a pseudo-field $\tilde{\bm{B}}$ and an applied field $\bm{B^{'}}$ measured from the sweet spot, yielding a qubit Hamiltonian $\mathcal{H}_{\rm q} =\gamma\bm{S} \cdot (\tilde{\bm{B}} +  \bm{B^{'}})$.
Without loss of generality, we assume coupling along the qubit's x-axis, leading to the interaction $\mathcal{H}_{\rm c} = \hbar g (\hat{a}^{\dagger} + \hat{a})\sigma_{x}$.
The orthogonality $\tilde{\bm{B}}\perp \bm{B^{'}}$ inherently introduces a transverse coupling component, manifesting experimentally as avoided level crossings (cf.~\crefadd{fig: Fluxon}{d}). The full AQRM thus reads:
\begin{equation}
    \mathcal{H}_{\rm AQRM}  = \mathcal{H}_\text{r} + \mathcal{H}_\text{c} +  \gamma\bm{S} \cdot \tilde{\bm{B}} + \gamma\bm{S} \cdot \bm{B^{'}}.
    \label{eq: general}
\end{equation}
We parametrize the fields via angles $\theta$ and $\phi$:
\begin{gather}
    \tilde{\bm{B}} = \tilde{B}
    \begin{pmatrix}
        \cos{\theta}\\
        0\\
        \sin{\theta}\\  
    \end{pmatrix},\\
    \bm{B}' = B'
    \begin{pmatrix}
        -\sin{\phi}\,\sin{\theta}\\
        \cos{\phi}\\
        \sin{\phi}\,\cos{\theta}\\  
    \end{pmatrix},
\end{gather}
as indicated in~\crefadd{fig:fitting}{a}.
To match the observed avoided crossings at symmetric values of $\bm{B}'$ around the sweet spot, the magnitude of the projection of the effective field $\bm{B}_{\rm eff} = \bm{B}' + \bm{\tilde{B}}$ onto the coupling axis $x$ must remain invariant under $\bm{B}' \rightarrow -\bm{B}'$.
This limits the model to $\phi=0$ for all $\theta$, except at $\theta = 0$ and $\theta=\pi/2$, where $\phi\in[0, \pi)$.

Among the resulting scenarios, we focus on three representative cases (cf.~\crefadd{fig:fitting}{b-d}), jointly fitting both the VQ and resonator frequency dependence (cf.~\crefadd{fig: Fluxon}{d,e}) in field.
All other configurations can be mapped onto these three.

In the configuration where the pseudo-field aligns with the coupling axis (cf.~\crefadd{fig:fitting}{b}), $\bm{B}'$ lies in the yz-plane.
For $\phi = 90^\circ$ the Hamiltonian becomes:
\begin{equation*}
    \mathcal{H}_{\{0^\circ , 90^\circ\}}  = \mathcal{H}_{\rm r} + \mathcal{H}_{\rm c} + \frac{\hbar\gamma}{2}\sigma_x \cdot \tilde{B} +  \frac{\hbar\gamma}{2}\sigma_z \cdot B'
\end{equation*}
This corresponds to pure longitudinal coupling at the sweet spot, resulting in zero dispersive shift~\cite{Richer2016Apr}, in contrast to the experimentally observed $\chi$.

\Crefadd{fig:fitting}{c} illustrates the case of pure transverse coupling, with both fields orthogonal to the coupling axis.
The Hamiltonian
\begin{equation*}
    \mathcal{H}_{\{90^\circ, 0^\circ\} }  = \mathcal{H}_{\rm r} + \mathcal{H}_{\rm c} + \frac{\hbar\gamma}{2}\sigma_z \cdot \tilde{B} +  \frac{\hbar\gamma}{2}\sigma_y \cdot B'
\end{equation*}
can be mapped to the SQRM (cf.~\cref{eq: SQRM}) and fails to capture the non-monotonic dependence of $\chi$ (cf. yellow line in~\crefadd{fig:fitting}{e}).

The best agreement with the measured dispersive shift is achieved in the configuration shown in~\crefadd{fig:fitting}{d}, where the applied magnetic field is aligned with the coupling axis. This introduces a longitudinal component in the interaction away from the sweet spot and gives the Hamiltonian:
\begin{equation*}
    \mathcal{H}_{\{90^\circ, 90^\circ\} }  = \mathcal{H}_{\rm r} + \mathcal{H}_{\rm c} + \frac{\hbar\gamma}{2}\sigma_z \cdot \tilde{B} -  \frac{\hbar\gamma}{2}\sigma_x \cdot B'\,,
\end{equation*}
identical to~\cref{eq: Hamiltonian} used to fit the data in the main text. 

\section{\label{app:hyperbolic_freq} Magnetic field Dispersion from a Vortex Tunneling Model}

To model VQ states we focus on the minimal configuration that yields strong anharmonicity: a double-well potential formed by two closely spaced pinning sites, allowing significant wavefunction delocalization.
The applied magnetic field modulates this potential through the Gibbs free energy of the vortex, which depends linearly on $B$ (cf.~\cref{eq:Gibbs}). 
A quantum mechanical model of vortex tunneling is developed to characterize the qubit's frequency, magnetic field dispersion, and anharmonicity.

The Gibbs free energy from \cref{eq:Gibbs} (cf.~\cite{Bronson_Gelfand_Field_2006, Kuit_2008}) is $\hat{G_1} = \hat{V}_\text{M}+\hat{V}_\text{SE}$, where the Meissner term $\hat{V}_\text{M}$ arises from the interaction between a Pearl vortex, at position ${x}$ along the wire width, and screening currents induced by a perpendicular magnetic field $B$. The term $\hat{V}_\text{SE}$, independent of magnetic field, gives the self energy due to the vortex current.
Assuming the vortex is far from the film edges along the wire length ($y$-direction),
\begin{equation}
    \hat{V}_\text{M} = 2\pi\varepsilon_0\left(\frac{B - n \Phi_0}{\Phi_0}\right) \hat{x} \left( \hat{x} - w \right), 
\end{equation}
where $n$ is the surface density of vortices, and 
\begin{equation}
    \hat{V}_\text{SE} =  \varepsilon_0\ln{ \left[ \frac{2 w}{\pi \xi}  \sin \left( \frac{\pi \hat{x}}{w} \right) +1\right]}.
\end{equation}
We model the two pinning sites using two-dimensional Lorentzian potentials:
\begin{equation}
    \hat{V}_\text{pin} = - \displaystyle\sum_{i=1,2} V_i \left(1 +  \frac{(\hat{x}-x_i)^2+(\hat{y}-y_i)^2}{\sigma_i^2}\right)^{-1}
    \label{eq: pinning_potential}
\end{equation}
where $V_i$, $\sigma_i$, and $(x_i, y_i)$ denote the depth, width, and position of each pinning site, respectively. 
Therefore the total potential energy is:
\begin{equation}
    \hat{V} =\hat{V}_\text{M} + \hat{V}_\text{SE} +\hat{V}_\text{pin}.
    \label{eq: pinning}
\end{equation}

Following ~\cref{eq: pinning}, the Hamiltonian for a pinned vortex of mass $m_{\rm v}$ is
\begin{equation}
    \mathcal{H} = -\frac{\hbar^2}{2 m_{\rm v}} {\nabla}^2 + \hat{V}(\hat{x},\hat{y};B).
    \label{eq:vortex_hamiltonian}
\end{equation}

\begin{figure}[tb]
    \includegraphics[width=0.95\columnwidth]{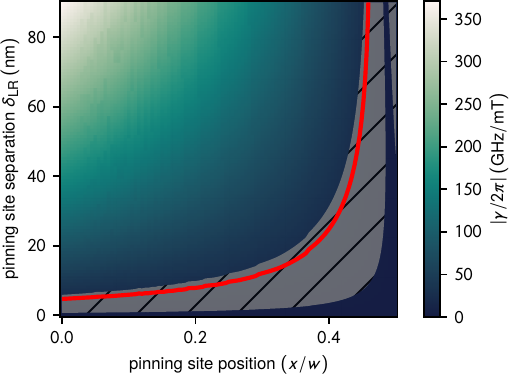}
    \caption{\label{fig:pinning} 
    \textbf{VQ magnetic field dispersion.}
    Calculated gyromagnetic ratio $\gamma$ (\cref{eq: Hamiltonian}) as a function of the separation $\delta_\mathrm{LR}$ between two pinning sites forming the double-well potential, plotted against the position of the first site along the resonator width. 
    Following~\cref{eq:calcGamma}, $\gamma$ quantifies the tunability of the double-well spectrum with external magnetic field $B$, i.e., the eccentricity of the hyperbolic qubit spectrum. 
    The hatched area indicates the range of $\gamma=\qtyrange{3}{25}{\giga\hertz/\milli\tesla}$ values extracted from Hamiltonian fits across all measured VQ–resonator systems (cf.~\crefadd{fig: correlations}{c}).
    The red line indicates the measured $\gamma = \qty{20}{\giga \hertz / \milli \tesla}$ corresponding to the VQ in main text. 
    It is important to note that although the Gibbs free energy favors a central position of the vortex ($x/w \approx 0.5$), in the presence of pinning its position can be significantly off-centered~\cite{Bai_2025}.
    }
\end{figure}

Due to the larger number of degrees of freedom of the microscopic model compared to the extracted parameters from the hyperbolic VQ spectrum, an unambiguous fit of the data to Hamiltonian~\cref{eq:vortex_hamiltonian} is not possible.
Nevertheless, from the measured $\gamma$, $B_0$ and $\omega_q(B_0)$, we can constrain the separation between pinning sites $\delta_{\rm LR}$, the pinning potential asymmetry $V_1 - V_2$, and we can provide an estimate for the zero point fluctuation amplitude of the vortex in its pinning sites $y_{\rm zpf}$, respectively. In the following, we discuss these quantities individually.

For a separation $\delta_\mathrm{LR}$ between sites, the Gibbs energy difference is
\begin{align}
    \Delta G(x,\delta_\mathrm{LR};B) &= |G_1(x - \delta_\mathrm{LR}/2 ,B) - G_1(x+\delta_\mathrm{LR}/2, B)|\nonumber\\
    &= |C(x, \delta_{\rm LR}) - h\gamma(x, \delta_{\rm LR})B|
\end{align}
where $x=(x_1+x_2)/2$ with $x_{1,2}$ the positions of the two wells along the width of the resonator, and $C$ is a constant independent of $B$.
We can identify 
\begin{align}
h\gamma = 2\pi\left(\frac{\varepsilon_0}{\Phi_0}\right)\big|\delta_{\rm LR}(2x - w)| ,
\label{eq:calcGamma}
\end{align}
as the gyromagnetic ratio in the AQRM (cf.~\cref{eq: Hamiltonian} and \cref{app:model}), resulting in a hyperbolic VQ spectrum, corroborating the single-vortex model for the VQ.

In \cref{fig:pinning}, the experimentally extracted range of $\gamma/2\pi=\qtyrange{3}{25}{\giga\hertz/\milli\tesla}$ (hatched region) indicates a large range of VQ locations where the pinning site separation is on the order of tens of nanometers, setting a microscopic length scale for the effective double-well potential, consistent with flux tunneling observed in granular aluminum nanojunction qubits (cf.~Refs.~\cite{Rieger_Gralmonium_2022, Günzler_2025}).
For each VQ, the measured $\gamma$ imposes a constraint on $x=(x_1+x_2)/2$ in relation to $\delta_{LR}=\big|x_1-x_2\big|$, as shown by the line in~\cref{fig:pinning}, corresponding to the VQ discussed in the main text. 

The measured $B_0$ imposes a second constraint, corresponding to the energy alignment of pinning sites $V(x_1,y_1;B_0)$ and $V(x_2,y_2;B_0)$. This takes the form
\begin{equation}
    \frac{\hbar}{2}\left(\Omega_{1}-\Omega_{2}\right) \approx \left(V_1-V_2\right)+\Delta G(x,\delta_\mathrm{LR};B_0),
    \label{eq:B_0_constraint}
\end{equation}
with $\Omega_{i}=\sqrt{k_i/m_{\rm v}}$ the frequency corresponding to the curvature of the pinning potentials $k_i=2V_i/\sigma_i^2$. For simplicity, we assume $k_1=k_2\equiv k$ and therefore, $\Omega_1=\Omega_2\equiv\Omega$, which gives
\begin{equation}
    V_2 \approx V_1+\Delta G(x,\delta_\mathrm{LR};B_0).
    \label{eq:B_0_constraint_V12}
\end{equation}
\Cref{eq:B_0_constraint_V12} sets the scale for the variability of pinning potential depths that give rise to VQs (cf.~\cref{app:Multiple_traps}).

Finally, fitting $\omega_q(B_0)$ imposes a constraint on the lowest two eigenenergies of Hamiltonian $\mathcal{H}(B_0)$,
\begin{equation}
    \mathcal{H}(B_0) = \hbar\Omega\left[-y_\text{zpf}^2 {\nabla}^2 + \frac{\hat{V}(\hat{x},\hat{y};B_0)}{\hbar\Omega}\right],
    \label{eq:H_B_0}
\end{equation}
where we have introduced the characteristic zero point fluctuation $y_\text{zpf}=\sqrt{\hbar/2m_{\rm v}\Omega} = ((8V_i m_{\rm v})^{0.25}/\sqrt{\hbar\sigma_i})^{-1}$. 
In principle, one could numerically solve the time-independent Schr\"odinger equation corresponding to \cref{eq:H_B_0} to fit $y_\text{zpf}$. However, this approach would require specific assumptions about the shape of the pinning potential, i.e., the microscopic pinning mechanism, which remains uncertain. While a promising direction for future research, this is beyond the scope of the present work. For now, we simply argue that to observe significant qubit frequencies in the \unit{\giga \hertz} range at $B_0$, $y_\text{zpf}$ must be in the range $\sigma \lesssim y_\text{zpf} \lesssim \delta_{\rm{LR}}$. As we will show in the next section, this estimate proves useful for evaluating the VQ-resonator coupling strength.

\section{\label{app:understanding}VQ-resonator Interaction}
We model the vortex as a cylinder of current traveling through the superfluid, similar to a whirlpool traveling in a liquid. The kinetic energy of superconducting charge carriers in the vortex is given by $E_\text{kin} = m|\bm{v}|^2/2$, where $m=2m_{\rm e}$ is the carrier mass and $\bm{v} = \bm{v}_{\rm r} + \bm{v}_{\rm v}$ combines the velocity induced by the resonator current, $\bm{v}_{\rm r}$, and the velocity of the moving vortex core, $\bm{v}_{\rm v}$.
In the absence of a normal vortex core, we model the kinetic energy of the vortex as the total kinetic energy of the $N$ charge carriers,
\begin{align*}
E_\text{kin} = N m_e|\bm{v}_{\rm r}+\bm{v}_{\rm v}|^2,
\end{align*}
with $N=n_{\rm s} S t$, where $n_{\rm s} = \frac{m_{\rm e}}{\mu_0 e^2 \lambda_L^2}$ is the carrier density, $S$ the effective vortex area and $t$ the film thickness. 
Assuming the resonator current $I_{\rm r}$ flows along the $y$-axis, the carrier velocity is $\bm{v}_{\rm r}=[I_{\rm r}/(q n_{\rm s} A)]\bm{e}_y$, where $q=2e$ is the carrier charge and $A=wt$ is the film's cross-sectional area.
The vortex velocity can be written as $\bm{v}_v={v}_{\mathrm{v},x}\bm{e}_x+{v}_{\mathrm{v},y}\bm{e}_y$, so that the kinetic energy becomes,
\begin{align*}
E_\text{kin} &= N m_e|({v}_{\rm r}+{v}_{\mathrm{v},y})\bm{e}_y+{v}_{\mathrm{v},x}\bm{e}_x|^2\\
&=N m_e({v}_{\rm r}+{v}_{\mathrm{v},y})^2+{v}^2_{\mathrm{v},x},
\end{align*}
and the resulting coupling term takes the form
\begin{align*}
H_\text{int} =2 N m_e v_{\rm r} {v}_{\mathrm{v},y} =\frac{S}{w} \frac{m_e}{e} I_{\rm r} {v}_{\mathrm{v},y}.
\end{align*}

We can express the current as $I_{\rm r}=\Phi \omega_{\rm r}/Z_{\rm r}$, with $Z_{\rm r}$ and $\omega_{\rm r}$ the resonator impedance and frequency, respectively. The flux operator can be expressed as $\hat{\Phi} = \Phi_\text{zpf}(\hat{a}^\dagger+\hat{a})$, where $\Phi_\text{zpf}=\sqrt{\hbar Z_{\rm r}/2}$. The quantized interaction becomes
\begin{equation}
\mathcal{H}_\text{int} = \frac{S}{w} \frac{m_e}{e} \frac{\omega_{\rm r}}{\sqrt{2 Z_{\rm r}/\hbar}}\: (\hat{a}^\dagger+\hat{a})\: \hat{v}_{\mathrm{v},y}.
\end{equation}
Within a factor of order one, depending on the orientation of the pinning sites along the resonator axis, ${v}_{\mathrm{v},y} \approx {v}_{\rm v}$ such that $(\hat{v}_{\mathrm{v},y}/{v}_{\rm v})\simeq\hat{\sigma}_y$ in the qubit basis. The interaction simplifies to
\begin{align}
\mathcal{H}_\text{int} &= \frac{S}{w} \frac{m_e}{e} \frac{\omega_{\rm r}}{\sqrt{2 Z_{\rm r}/\hbar}}\: (\hat{a}^\dagger+\hat{a})\: \: \hat{\sigma}_y {v}_{\mathrm{v},y}\\
&= \tilde{g}\: (\hat{a}^\dagger+\hat{a})\: \hat{\sigma}_y,
\end{align}
with coupling strength
\begin{align}
    \tilde{g} &=\frac{S}{w} \frac{m_e}{e} \frac{\omega_{\rm r}}{\sqrt{2 Z_{\rm r}/\hbar}}\: {v}_{\mathrm{v},y}.
\end{align}

Approximating the bottom of the pinning potential (cf.~\cref{app:hyperbolic_freq}) to be harmonic, the magnitude of vortex velocity is given by ${v}_{\rm v}=\hbar/(m_{\rm v} y_\text{zpf})$, where $m_{\rm v}$ is the total mass of the vortex and $y_\text{zpf}^{-1}=(8V_i m_{\rm v})^{0.25}/\sqrt{\hbar\sigma_i}$ is the characteristic zero point fluctuation length. 
Using the von Klitzing constant $R_{\rm K}=h/e^2= \qty{25.8}{\kilo \ohm}$ and substituting $y_{\rm zpf}$, the coupling strength becomes,
 \begin{align}
    \tilde{g} &=\hbar\omega_{\rm r}\: \frac{S}{w}\frac{m_{\rm e}}{m_{\rm v}} \frac{1}{y_\text{zpf}}
    \sqrt{\frac{R_{\rm K}}{4 \pi Z_{\rm r}}},
\end{align}
in which we recognize $m_{\rm v}/m_{\rm e}=N=n_{\rm s}St$. After substitution and simplification, we obtain
\begin{align}
    \tilde{g}
    &= \hbar\omega_{\rm r}\:\frac{1}{wt}\frac{\lambda_L^2}{y_\text{zpf}} \frac{\mu_0 e^2}{m_{\rm e}} \sqrt{\frac{R_{\rm K}}{4\pi Z_{\rm r}}}.
\end{align}
For $Z_r\simeq \qty{3}{\kilo \ohm}$, $\lambda_{\rm L} \approx \qty{4}{\micro \meter}$, and the zero point fluctuations estimated in \cref{app:hyperbolic_freq}, $\qty{1}{\nano \meter} \lesssim y_\text{zpf} \lesssim \qty{10}{\nano \meter}$, we find $\tilde{g}/\omega_{\rm r} \simeq \qtyrange{0.1}{1}{\percent}$, consistent with the measured range of coupling strengths reported in~\crefadd{fig: correlations}{d}. 

\section{\label{app:vortex-vortex}Vortex-Vortex Interaction}
\begin{figure}[tb]
    \includegraphics[width=1.0\columnwidth]{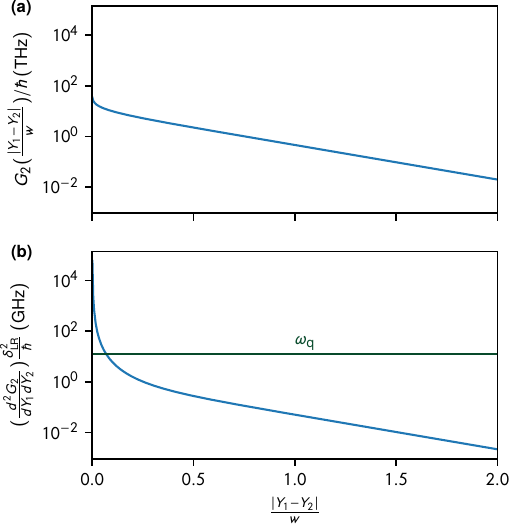}
    \caption{\textbf{Interaction between vortices.}
        \textbf{(a)} Mutual interaction potential $G_2$ between two vortices located along the center of the wire ($X_1=X_2=w/2$) as a function of their separation $|Y_1-Y_2|$. 
        \textbf{(b)} Estimated qubit–qubit interaction strength between two vortex-qubits pinned at positions $X_1=X_2=w/2$, versus their separation along the length of the resonator $|Y_1-Y_2|$, for $\delta_{\rm {LR}} = \qty{10}{\nano\meter}$. 
        The green line indicates one of the measured qubit frequencies $\omega_{\rm q}$.
        }
    \label{fig:interaction}
\end{figure}

Field cooling above $\phi \gtrsim \phi_{\rm s}$ induces a large number of vortices, which may become trapped at various pinning sites along the wire. 
To assess the impact of vortex-vortex interaction on the dynamics of the vortex-qubit, we derive the interaction Hamiltonian $\mathcal{H}_\mathrm{int}$ between two pinned vortices.
We begin with the Gibbs free energy due to interaction between two Pearl vortices ($i=\{1,2\}$), located at positions $\bm{r}_i =(x_i,y_i)$  \cite{Bronson_Gelfand_Field_2006}:
\begin{align}
G_2(\bm{r}_1,\bm{r}_2)= \varepsilon_0 \ln \left(\frac{\cosh (\tfrac{\pi(y_1 - y_2)}{w})-\cos (\tfrac{\pi (x_1 + x_2)}{w})}{\cosh (\tfrac{\pi(y_1 - y_2)}{w})-\cos (\tfrac{\pi(x_1 - x_2)}{w})}\right)
\end{align}
where $\varepsilon_0 = \Phi_0^2/(2\pi\mu_0\Lambda)$. 
Here, the wire geometry is defined over $y \in[0, L]$ along its length and $x \in[0,w]$ along its width. 
This interaction is repulsive and decays exponentially over length scales on the order of $w$, as shown in \crefadd{fig:interaction}{a}.

We assume each vortex is trapped at a pair of pinning sites, with its center located at $\bm{R}_i(X_i,Y_i)$. 
We define local vortex coordinates $\bm{r}_i=(x_i,y_i)$, measured relative to the pinning site center. 
In the local qubit basis, the position operator can be expressed as:
\[ \hat{r}_i^{(\xi)}=\delta_{\rm LR}\left(\alpha_i^{(\xi)}\hat{\sigma}_x+\beta_i^{(\xi)}\hat{\sigma}_z\right)\] 
where $\xi\in \{x,y\}$, $\delta_{\rm LR}$ is the length scale of vortex tunneling, and the coefficients $\alpha_i^{(\xi)}$ and $\beta_i^{(\xi)}$ are dependent on the pinning site geometry, dimensionless, and typically of the order unity.

Assuming the vortex separation is large compared to the tunneling distance, $|\bm{R}_1-\bm{R}_2|\gg \delta_{\rm LR}$, we linearize the interaction to obtain the effective Hamiltonian:
\begin{align}
\mathcal{H}_\text{int}(\bm{r}_1,\bm{r}_2)=\hat{r}_1^{\:T}{\left[\left(\bm{\nabla}_{\bm{R}_1}\bm{\nabla}_{\bm{R}_2}\right)G_2(\bm{R}_1,\bm{R}_2)\right]\:\hat{r}_2}, 
\end{align}
where $\left(\bm{\nabla}_{\bm{R}_1}\bm{\nabla}_{\bm{R}_2}\right)G_2(\bm{R}_1,\bm{R}_2)$ is the $2\times2$ mixed-derivatives Hessian matrix of $G_2(\bm{R}_1,\bm{R}_2)$, defined as:
\[
\left[\left(\bm{\nabla}_{\bm{R}_1}\bm{\nabla}_{\bm{R}_2}\right)G_2(\bm{R}_1,\bm{R}_2)\right]_{\xi,\xi^\prime} = \frac{\partial^2 G_2}{\partial R_1^{(\xi)}\partial R_2^{(\xi^\prime)}},
\]
for $\{\xi,\xi^\prime\}\in\{x,y\}$.
To estimate the typical qubit-qubit coupling strength, we consider a simplified geometry where both vortex centers lie along the central axis of the wire and are aligned along the $y$-direction.
Under this condition, the only non-zero component of the interaction tensor is $\left[\left(\bm{\nabla}_{\bm{R}_1}\bm{\nabla}_{\bm{R}_2}\right)G_2(\bm{R}_1,\bm{R}_2)\right]_{y,y}$. 
The resulting interaction strength is plotted in \crefadd{fig:interaction}{b} and compared to a representative qubit frequency $\omega_q = \qty{12.5}{\giga \hertz}$.

Due to the low vortex density generated during field cooling ($ N_\odot\simeq 1$), it is unlikely that two vortices become trapped at separations significantly smaller than $w$. 
In this case, the vortex-vortex interaction strength is at least one order of magnitude smaller than the typical vortex-qubit frequency (cf.~\cref{fig:interaction}).
Moreover, since $\omega_{\rm q}$ depends exponentially on the tunneling barrier, small variations in local pinning geometries can lead to detunings comparable to or larger than the qubit frequency.
This further suppresses coherent coupling between neighboring vortex qubits. 
These facts support the interpretation that a single pinned vortex accounts for the experimental observations presented in the main text.

\section{\label{app:more_info} Extended Time Domain Characterization and Quantum Jumps of VQs}
\begin{figure*}[t]
 \begin{minipage}[t]{1.0\columnwidth}
    \includegraphics{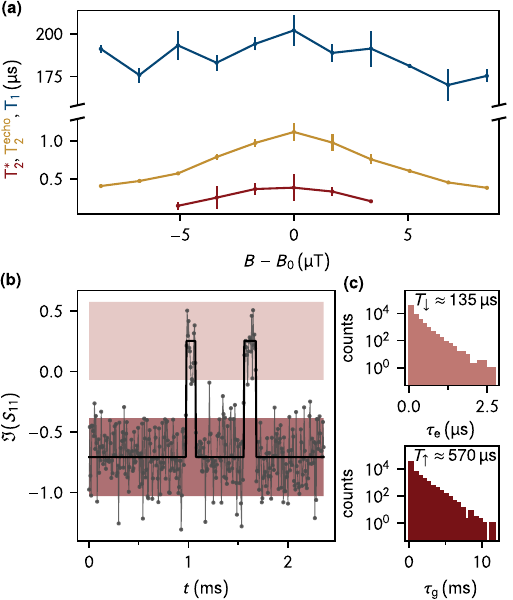}
    \caption{\label{fig:QJ}\textbf{Extended time domain characterization for the main text VQ.}
    \textbf{(a)}~Energy relaxation time $T_1$ (blue), Ramsey coherence time $T_2^*$ (red), and Hahn echo coherence times $T_2^{\mathrm{echo}}$ (yellow) as a function of the applied magnetic field detuning from the sweet spot $B-B_0$. 
    Markers and error bars denote the mean and standard error of the mean of two consecutive field sweeps, each averaging \num{5000} single-shot measurements.
    \textbf{(b,c)}~Quantum jump detection of the VQ at the sweet spot $B_0$.
    \textbf{(b)}~Representative time trace of the imaginary part of the resonator reflection coefficient $\Im(S_{11})$ (gray markers).
    Qubit states are assigned using a latching filter (black line) based on $1.5$ standard deviation thresholds (dark red band for $\ket{g}$, light red for $\ket{e}$) centered on their respective means. 
    \textbf{(c)}~Histograms of dwell times in the ground (lower panel) and excited (upper panel) states from \num{5e6} single-shot measurements. 
    Average excitation and relaxation times are $T_{\uparrow} = \qty{570}{\micro \second}$ and $T_{\downarrow} = \qty{135}{\micro \second}$ respectively.
    }
  \end{minipage}
  \hspace{0.43cm}
 \begin{minipage}[t]{1.0\columnwidth}
    \includegraphics{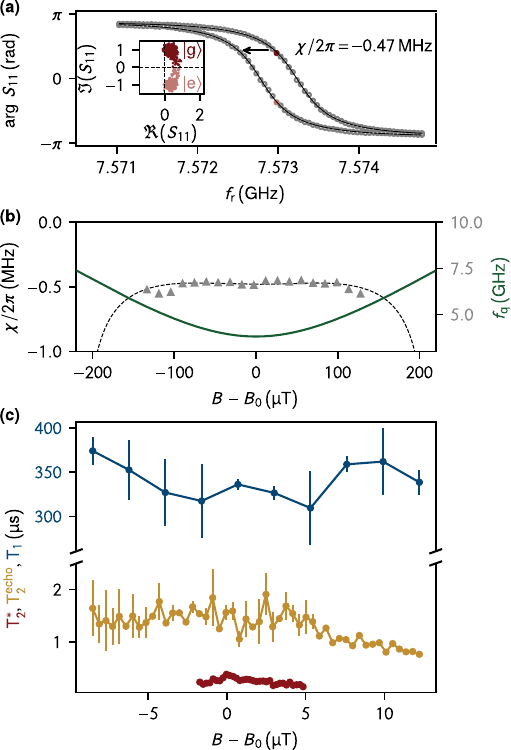}
    \caption{\label{Fig: another_trap}\textbf{Characterization of a VQ introduced in a ZFC sample.} The VQ is introduced by ZFC of the grAl resonator followed by ramping to \qty{1}{\milli\tesla} (cf.~\crefadd{fig: Alltraps}{c}). \textbf{(a)}~Dispersive shift measurement (similar to \crefadd{fig: TD_Chi}{b}), yielding $\chi = \qty{-0.47}{\mega\hertz}$. 
    Dark red ($\ket{g}$) and light red ($\ket{e}$) markers denote data extracted from the IQ clouds shown in the inset. 
    \textbf{Inset:}~Normalized scatter plot of repeated $S_{11}$ measurements in the complex plane, referenced to the ground state amplitude.
    Note that additional IQ cloud histograms for this VQ are presented in~\cref{fig: Dispersive}.
    \textbf{(b)} Measured dispersive shift $\chi$ (triangles) near $B_0$ (similar to \crefadd{fig: TD_Chi}{d}). 
    The green line is a fit of the qubit spectrum (right axis) to \cref{eq: Hamiltonian} while the dashed line indicates the predicted $\chi$-dependence. 
    \textbf{(c)} Extracted energy relaxation time $T_1$ (blue), and coherence times $T_2^*$ (red) and $T_2^{\mathrm{echo}}$ (yellow) as a function of detuning from $B_0$ (similar to \crefadd{fig:QJ}{a}). 
    Markers and error bars represent the mean and standard deviation of two consecutive field sweeps.
    }
   \end{minipage}
\end{figure*}

In \crefadd{fig:QJ}{a}, we present energy relaxation and coherence of the VQ as a function of magnetic field detuning from the sweet spot.
Whereas $T_1$ remains largely insensitive to detuning, coherence times decrease away from $B_0$, consistent with flux-noise-induced dephasing observed in superconducting flux qubits~\cite{Yoshihara2006Oct, stern2014flux, Yan_2016, Rieger_Gralmonium_2022}.
In \crefadd{fig:QJ}{b}, we use time-resolved $S_{11}$ measurements to detect quantum jumps. 
We apply \qty{1.2}{\micro\second} long readout pulses, spaced \qty{5}{\micro\second} apart and corresponding to an average photon number $\overline{n}_{\rm r} = 7$, similar to the main-text  measurements (cf.~\cref{fig: Times}).
State transitions are identified with a two-point latching filter, which registers a jump when $\Im(S_{11})$ enters the 1.5 $\times$ standard deviation band centered on the mean of the $\ket{g}$ or $\ket{e}$-states, as indicated by the shaded bands.
Dwell-time histograms (\crefadd{fig:QJ}{c}) give average excitation and relaxation times of $T_{\uparrow} = \qty{570}{\micro \second}$ and $T_{\downarrow} = \qty{135}{\micro \second}$, corresponding to a $T_1$ time of $T_1 = (T_{\downarrow}^{-1} + T_{\uparrow}^{-1})^{-1} = \qty{110}{\micro\second}$, within the temporal fluctuations of the free decay time shown in the main text~\crefadd{fig: Times}{a}.

\Cref{Fig: another_trap} provides a complete characterization of a VQ introduced after ZFC, yielding results consistent with those of the main-text sample. 
We measure a dispersive shift for this VQS-resonator system of $\chi/2\pi=\qty{0.5}{\mega \hertz} < \kappa/2\pi=\qty{0.75}{\mega \hertz}$ with a magnetic field dependence away from the sweet spot that aligns with the AQRM model (\crefadd{Fig: another_trap}{b}).
Moreover, we observe energy relaxation and coherence times (\crefadd{Fig: another_trap}{c}) comparable to those in FC VQs.
  
\end{document}